\documentclass[fleqn,10pt]{wlscirep}
\usepackage[utf8]{inputenc}
\usepackage[T1]{fontenc}
\usepackage{bm}
\usepackage{hyperref}
\usepackage{xurl}
\usepackage{newunicodechar}
\newunicodechar{ć}{\'c}
\usepackage{etoolbox}
\usepackage{multirow}
\usepackage{adjustbox}
\usepackage{subcaption}
\usepackage{float}
\usepackage{cleveref}
\usepackage{xcolor}

\usepackage[
backend=biber,
url=true,
style=nature,
sorting=none,
giveninits=true,    
maxnames=10,
autocite=inline   
]{biblatex}

\AtEveryBibitem{%
 \clearfield{note}   
   \ifentrytype{online}
    {}
    {\clearfield{url}}%
 \clearfield{issn}   
 \clearfield{isbn}   
 \clearlist{language}  
 \clearlist{urldate}
   \clearfield{urlurldate}%
  \clearfield{urlyear}%
  \clearfield{urlmonth}%
    \clearfield{month}%
  \clearfield{day}%
  \clearfield{endmonth}%
  \clearfield{endday}%
  \clearfield{eventmonth}%
  \clearfield{eventday}%
}

\title{Recurrent visitations reveal selectivity beyond the 15-Minute City vision}

\author[1,*]{Xiuning Zhang}
\author[2]{Alexei Poliakov}
\author[3]{Henrikki Tenkanen}
\author[1]{Elsa Arcaute}

\affil[1]{The Centre for Advanced Spatial Analysis, University College London, London, UK}
\affil[2]{Locomizer Ltd, London, UK}
\affil[3]{Department of Built Environment, Aalto University, Espoo, Finland}

\affil[*]{e-mail: xiuning.zhang.23@ucl.ac.uk}

\begin{abstract}
In the transition towards sustainable and equitable urban living, proximity-centred planning has been adopted in cities worldwide. Exemplified by the 15-Minute City (15mC), this planning paradigm often assumes that local amenity provision translates into local use, yet behavioural evidence on recurrent visitation remains limited. 
To address this gap, we introduce $K$-Visitation, a scalable, behaviourally informed framework for comparing recurrent visitation with proximity-based expectations, using 18 months of mobile phone data from 720,000 users in Finland. For each individual, $K$-Visitation compares recurrent destinations ($K_{freq}$) with nearest options ($K_{dist}$), 
and evaluates travel time differences between them and their neighbourhood proximity baseline. Across Finnish cities, individuals repeatedly bypassed nearby visited options that already exposed them to required daily amenity categories, incurring measurable travel time costs beyond what local opportunities could support. The sharpest divergence appears in amenity-rich urban cores: where the physical conditions for 15-minute living are strongest, recurrent destinations depart most from nearest options.
A density-aware null model and destination-level classifier show that this divergence reflects selective behaviour: non-nearest recurrent destinations are not explained by amenity density alone and are often amenity-rich but functionally specialised. Amenity-specific analysis further reveals a hierarchy in which routine anchors adhere more closely to proximity, while specialised and infrastructural functions depend on wider catchments and connectivity.
Our findings show that proximity is a necessary but insufficient condition for local living. 
The 15mC should therefore secure routine local access while using behavioural evidence to identify where proximity is bypassed and where wider connectivity remains necessary.
\end{abstract}

\begin{document}

\begin{refsection}
\flushbottom
\maketitle

\thispagestyle{empty}

Cities are navigating intertwined challenges--from the imperative to achieve net-zero emissions to the socio-economic aftershocks of the pandemic \autocite{un-habitat2020people}. In response, cities are pivoting away from car-centric development, and moving towards locally oriented models of urban life \autocite{un-habitat2020people, c40202115minute}. Among the ideas gaining momentum, the 15-Minute City (15mC) has emerged as a prominent framework: residents should be able to access essential services, such as groceries, schools, healthcare, within a short distance from home \autocite{moreno2021introducing, allam202215minute}. Rooted in longer traditions of mixed-use, human-scale neighbourhoods and theories of urban vitality \autocite{perry1929neighborhood, jacobs1964death, kissfazekas2022circle, khavarian-garmsir202315minute}, the 15mC has achieved wide policy uptake and international diffusion \autocite{teixeira2024classifyinga}.

The significant role of local travel is also empirically documented within the mobility science literature. Individuals' repeated visitations follow distance-frequency trade-offs and scale-free patterns \autocite{gonzalez2008understanding, alessandretti2020scales, schlapfer2021universal}. The regularity of shorter distance travel underscores the importance of local options for everyday life, and adds support for the proximity-based planning paradigm.
While the 15mC offers a compelling narrative for sustainable transition, its reliance on a uniform temporal threshold (e.g., 15 minutes) risks oversimplifying the profound heterogeneity inherent in urban systems \autocite{mouratidis2024time, khavarian-garmsir202315minute}. 
The premise that a ``one-size-fits-all'' radius can serve diverse populations ignores critical disparities in urban form, transport infrastructure, and socioeconomic composition that render such thresholds inequitable or unrealistic in many contexts \autocite{aparicio2024walkability, bruno2024universal}. Most crucially, this decentralising vision disregards the basic hierarchy of urban centres and services \autocite{mouratidis2024time, birkenfeld2023who}, creating a one-size-fits-all target for different amenities to be localised; it overlooks foundational principles like Central Place Theory \autocite{christaller1933zentralen}, which posits that specialised, high-order amenities inherently depend on larger catchment areas and agglomeration economies that localised planning cannot replicate.
Furthermore, the 15mC concept posits that improved local amenity provision increases local trip frequency, with density and diversity as key ingredients for transforming the physical environment \autocite{moreno2021introducing}.
Consequently, current assessments predominantly rely on primal accessibility metrics, such as isochrones, to quantify the volume of opportunities reachable within a specific time \autocite{cui2020primal}. While essential for setting supply-side goals, this approach abstracts the complex utility functions within individual travel behaviour.
Although individuals are anchored by proximity \autocite{barbosa2018human, gonzalez2008understanding}, various factors, such as commuting and personal preferences, can lead to bypassing local options in favour of destinations further afield \autocite{manaugh2012what,ellder2022when, naess2005residential, naess2018causality}. 
As a result, a persistent gap emerges between the metrics used to support local-living scenarios and the behavioural reality of residents. Empirical studies show that even when accessibility metrics indicate a complete 15-minute neighbourhood, individuals often continue to travel beyond it \autocite{maciejewska2025when, birkenfeld2023who, abbiasov202415minute}.

Recent empirical efforts using large-scale human mobility data have begun to bridge this gap. Notably, Abbiasov et al. \autocite{abbiasov202415minute} used GPS data from the US to quantify ``15-minute usage,'' revealing that local trips constitute only a small fraction of daily consumption, despite being largely explained by local accessibility.
However, it remains unclear whether patterns observed in the car-dependent and low-density fabric of US cities also translate to dense, transit-oriented contexts such as European cities. Furthermore, neighbourhood-level trip shares cannot reveal whether proximity structures recurrent visitations, the stable and limited set of destinations that dominate daily life and anchor individual mobility \autocite{alessandretti2018evidence,gonzalez2008understanding,mazzoli2019field}.

In this paper, we introduce the $K$-Visitation framework, a behaviourally grounded method that compares the structural and temporal patterns of recurrent visitation against a proximity-based expectation. 
Using an 18-month dataset of anonymised mobility traces from 720 thousand users in Finland, we identify the set of visitations required to cover essential daily functions under two distinct schemes: one corresponding to frequency of visitations ($K_{freq}$) and the other to the observed visitations that are closer to home ($K_{dist}$).
We apply this framework to investigate the central research question: \textit{To what extent does recurrent visitation align with the proximity-based expectations of the 15mC, and what explains their alignment or divergence?}

Our central finding is that recurrent visitation systematically departs from the nearest options that would already satisfy residents' amenity needs: individuals repeatedly travel further and bear a measurable travel time cost to do so. Fulfilling amenity categories through proximity is therefore necessary but not sufficient to explain how people actually move, a distinction that the supply-side framing of the 15mC has largely overlooked.

In doing so, this study advances the literature on accessibility and the 15mC through three contributions.
First, methodologically, we introduce the $K$-Visitation framework to compare recurrent visitation with proximity-based opportunities. By distinguishing spatial alignment from travel time differences, the framework complements conventional supply-side accessibility measures with a behavioural perspective on local living.
Second, empirically, we examine how the relationship between proximity-based accessibility and recurrent visitation varies across cities and neighbourhood contexts. 
Third, conceptually, we investigate how destination characteristics and amenity functions are associated with recurrent visitations beyond the nearest options. This provides a basis for understanding when proximity may anchor everyday mobility and for place-sensitive development of the 15mC.
\section*{Results}

\subsection*{K-Visitation framework}

Leveraging large-scale mobile phone GPS data in Finland, we introduced the \emph{K-Visitation} framework to evaluate how closely individual mobility aligns with this proposition. 

We first define each visitation as exposure to amenities. Each time an individual stays at a place, they are exposed to the amenity categories present there (see ``Amenity and social exposure through visitation'' in Methods). From the 15mC literature we derive ten daily amenity categories, and for each individual we define their required categories as those they are exposed to at least once across the observed visited places.

\begin{figure}
    \centering
    \includegraphics[width=1\linewidth]{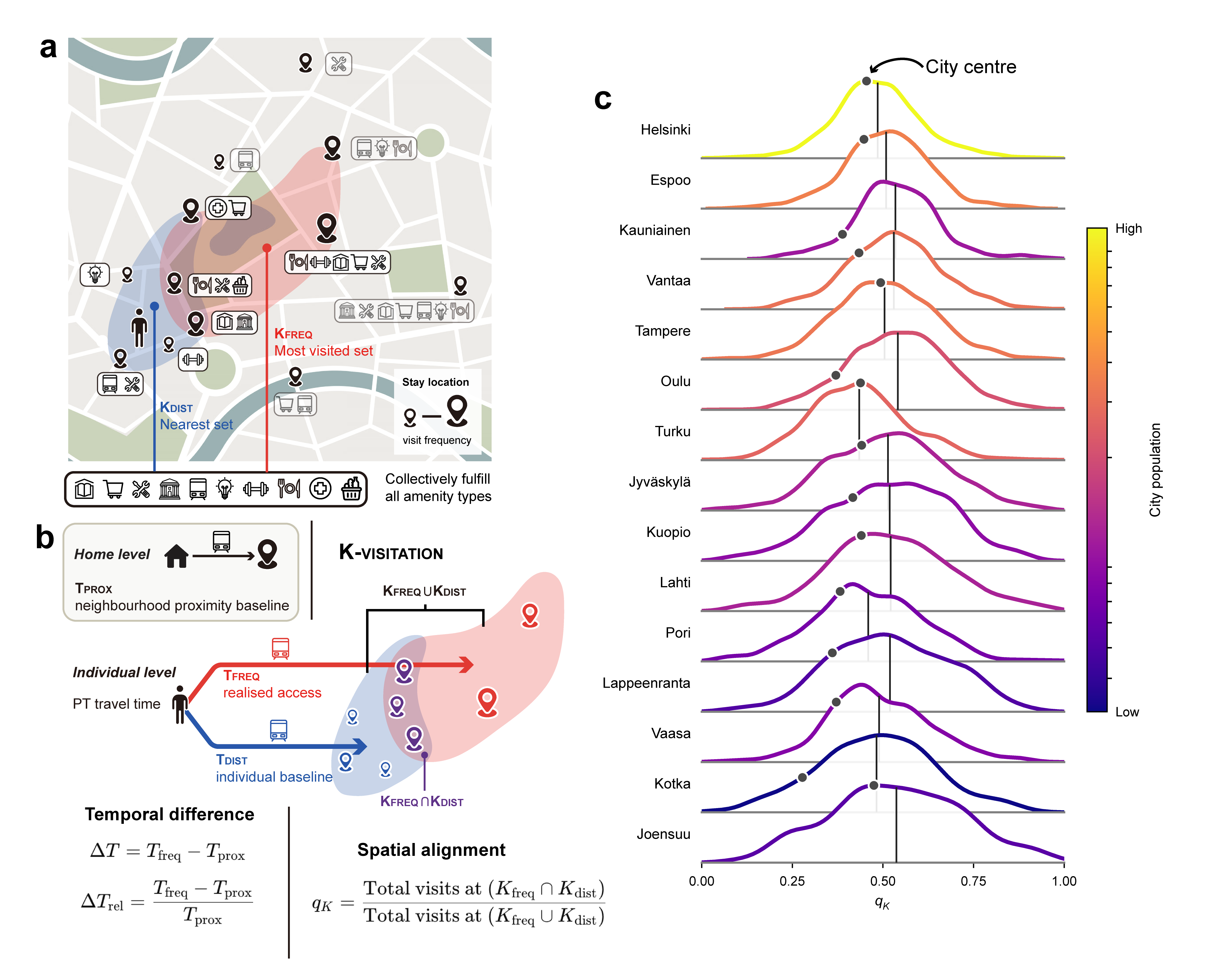}
    \caption{\textbf{K-Visitation framework and spatial alignment across Finnish cities.} 
    \textbf{a.} Schematic of the framework: The recurrent visitation set $K_{freq}$ (red) and the nearest-location set $K_{dist}$ (blue); each provides exposure to all required amenity categories. Larger location markers indicate higher visit frequency, and icons represent exposure to different amenity types at each visitation.
    \textbf{b.} Temporal comparison among public transport travel time to recurrent destinations ($T_{freq}$), travel time to the nearest visited set ($T_{dist}$), and the neighbourhood-level proximity baseline ($T_{prox}$). The absolute and relative temporal differences compare $T_{freq}$ with $T_{prox}$ and are defined as $\Delta T=T_{freq}-T_{prox}$ and $\Delta T_{rel}=(T_{freq}-T_{prox})/T_{prox}$, respectively.
    Spatial alignment ($q_K$) measures the visit-frequency-weighted overlap between ($K_{freq}$) and $(K_{dist}$).
    \textbf{c.} Distribution of $q_K$ across 15 Finnish cities, coloured by city population (colour bar, low to high). Black dots denote the $q_K$ of the city centre in each city, and vertical lines represent the city-level mean. Cities are arranged in decreasing order of population (except for Espoo, Kauniainen and Vantaa, as they are part of the Helsinki metropolitan area), showing that larger cities do not exhibit higher alignment than smaller cities.
    }
    \label{fig:framework}
\end{figure}

The framework then asks a simple question: \emph{Does the subset of places an individual frequently visits to satisfy all required amenity categories correspond to their nearest options, or do they travel farther to reach alternative destinations?} To answer this question, we identify the subset of visited places, denoted $K$, that collectively covers the individual's required amenity categories.
From the individual's visitations (excluding home), places are selected iteratively in a specified order until all required categories are covered for the first time, at which point the scan stops. We apply this identical procedure under two different orderings: frequency and proximity to home (Fig.~\ref{fig:framework}a).

The \emph{recurrent set}, $K_{freq}$, orders places from most to least frequently visited, capturing the destinations that anchor an individual's routine, consistent with prior treatments of recurrent visitation in mobility science \autocite{gonzalez2008understanding,alessandretti2018evidence,mazzoli2019field}. 
The \emph{nearest set}, $K_{dist}$, orders the same visited places from nearest to farthest by Euclidean distance from home, so it represents the nearest options that the person did reach. The two sets impose identical category requirements and the same stopping rule, differing only in the order in which places are considered. A full step-by-step construction for a hypothetical individual is provided in Supplementary Section~S1.

We compare the two aspects of mobility using spatial alignment and travel time difference (Fig.~\ref{fig:framework}b).
First, spatial alignment is measured by $q_K$, the frequency-weighted overlap between $K_{freq}$ and $K_{dist}$:
\begin{equation}
q_K =
\frac{\text{Total visits at } (K_{freq}\cap K_{dist})}
{\text{Total visits at } (K_{freq}\cup K_{dist})},
\end{equation}
which is the frequency-weighted overlap between the two sets. Because $K_{dist}$ already satisfies amenity categories using the nearest available places, each individual could in principle meet all of their amenity needs through $K_{dist}$; $q_K$ therefore measures how far their recurrent choices depart from that possibility. A high $q_K$ (where $1$ represents perfect overlap) indicates that recurrent destinations coincide with the nearest options, whereas a low $q_K$ indicates that the individual repeatedly bypasses adequate nearby places---a revealed preference for further locations.

Second, we calculate the absolute and relative travel time differences using public transport, comparing individual-level travel times to $K_{freq}$ (defined as $T_{freq}$) with the neighbourhood-level proximity baseline $T_{prox}$ (see ``Travel time by public transport'' in Methods) as
\begin{equation}
\Delta T = T_{freq}-T_{prox},
\qquad
\Delta T_{rel} =
\frac{T_{freq}-T_{prox}}{T_{prox}}.
\end{equation}
Larger $\Delta T$ means that reaching the individual's recurrent destinations takes longer than what the nearby available opportunities would allow. The absolute difference, $\Delta T$, is expressed in minutes, while the relative difference, $\Delta T_{rel}$, is in proportion to the local benchmark.

Based on the 15mC premise, we expected residents of larger cities and urban centres, with denser amenities and higher public transport connectivity, to show greater spatial alignment (higher \(q_K\)) and smaller absolute and relative travel time differences.

\subsection*{Recurrent visitations diverge from nearest options}

\begin{figure}
  \centering
    \includegraphics[width=1\linewidth]{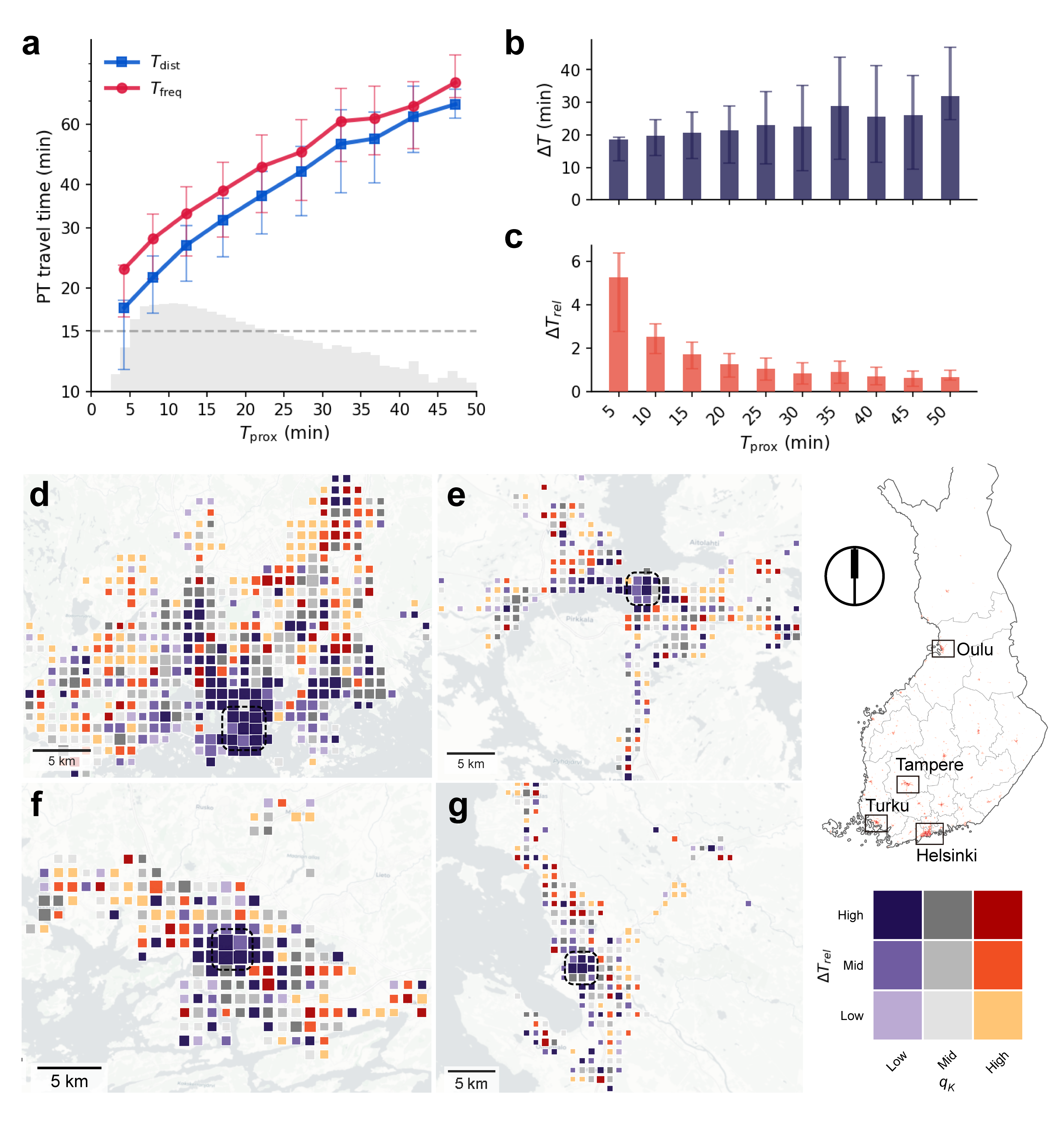}
    \caption{\textbf{Spatial and temporal divergence of recurrent visitation from proximity.}
    \textbf{a.} Mean public transport travel times to $K_{dist}$ (blue squares) and $K_{freq}$ (red circles) destinations, grouped by neighbourhood proximity time ($T_{prox}$). The grey distribution shows $T_{prox}$, and the dashed line marks 15~min.
    \textbf{b.} Mean absolute temporal difference, $\Delta T=T_{freq}-T_{prox}$, across $T_{prox}$ groups; the absolute difference widens towards poorly accessible areas. \textbf{c.} Mean relative temporal difference, $\Delta T_{rel}=(T_{freq}-T_{prox})/T_{prox}$, across $T_{prox}$ groups; the relative difference is greatest in highly accessible areas. Markers show means and error bars span the 25th--75th percentiles.
    \textbf{d--g.} Bivariate spatial distributions of $q_K$ and $\Delta T_{rel}$ across grid cells measuring $1~\mathrm{km}\times1~\mathrm{km}$ in \textbf{d.} Helsinki, \textbf{e.} Tampere, \textbf{f.} Turku and \textbf{g.} Oulu. Grid colours classify both measures as low, intermediate or high; cell size indicates POI density.
    Urban cores combine low spatial alignment with high relative temporal differences, whereas many peripheral areas show greater alignment with nearby destinations.
    }
    \label{fig:tt-bivar}
\end{figure}

\paragraph{Systematic divergence from nearest options}
$q_K$ in Finnish cities centres around 0.5, indicating that recurrent visitations form a significant part of local living, and this pattern is consistent across Finnish cities. Helsinki, the largest and most amenity-dense region, showed marginally lower $q_K$ than smaller cities, and Turku, the third-largest city, recorded the lowest mean alignment of all (Fig.~\ref{fig:framework}c).
Furthermore, the city centres, here defined as the area with the most POIs, exhibited systematically lower alignment than other areas. Mapping $q_K$ at the grid level showed that high-density city centres exhibited the strongest deviation from the baseline, whereas amenity-sparse peripheries showed the highest alignment (Supplementary Figs.~S9 and S10).
These results reveal an inversion of the expected spatial patterns: recurrent destinations align less with nearest options where local opportunities were most abundant, but align more closely where opportunities were sparse. 
The temporal analysis revealed a persistent travel time difference across the full accessibility gradient (Fig.~\ref{fig:tt-bivar}a--c). 
More accessible areas also showed shorter absolute travel times to recurrent destinations (Fig.~\ref{fig:tt-bivar}a).
Even in the most accessible areas where the required amenity categories could be reached within approximately 5--10~min under the proximity baseline, travel to recurrent destinations averaged about 25~min.
The relative difference, as a result, was greatest in these areas: residents travelled several times beyond what their local environment made possible. 

The joint spatial and temporal distribution revealed two contrasting mobility regimes (Fig.~\ref{fig:tt-bivar}d--g; Supplementary Fig.~S12). In connected, amenity-rich urban centres, low spatial alignment coincided with a high relative temporal difference. Residents used abundant local accessibility as a platform for reaching further recurrent destinations. Proximity therefore did not confine mobility; its greatest availability coincided with the strongest proportional departure from it. The 15mC premise that local abundance anchors local mobility thus broke down precisely where the physical conditions for proximity-based living were most fully met.
Poorly accessible peripheral areas exhibited the reverse pattern. Recurrent portfolios aligned more closely with nearby alternatives, but both the proximity baseline and realised travel involved substantially longer journeys. Their apparent localism may therefore reflect constraints with limited local options. This contrast is in accordance with existing travel behaviour literature \autocite{ellder2022when,naess2018causality} and challenges the assumption that greater local provision necessarily produces more local behaviour.

\paragraph{Individual mobility behaviour resists the pull of density}
The preceding results show that abundant accessibility can enable residents to bypass their nearest options. Yet this pattern raises an alternative explanation: the observed central pattern may simply reflect an attraction towards dense, amenity-rich destinations with distance decay. In this sense, the 15mC concept provides a workable target: through densification, local provision can be made sufficient to anchor recurrent visitation. 

We tested this explanation by comparing empirical alignment, $q_K$, with $q_K^{null}$ generated by a density-aware Exploration and Preferential Return (d-EPR) model \autocite{pappalardo2015returners} (see ``d-EPR null model'' in Methods). The model preserves generic mobility mechanisms while replacing observed destination choices with probabilistic attraction to amenity density constrained by distance. Thus, $q_K^{null}$ represents the alignment expected if recurrent visitation were structured by density. 

\begin{figure}
  \centering
  \includegraphics[width=0.7\linewidth]{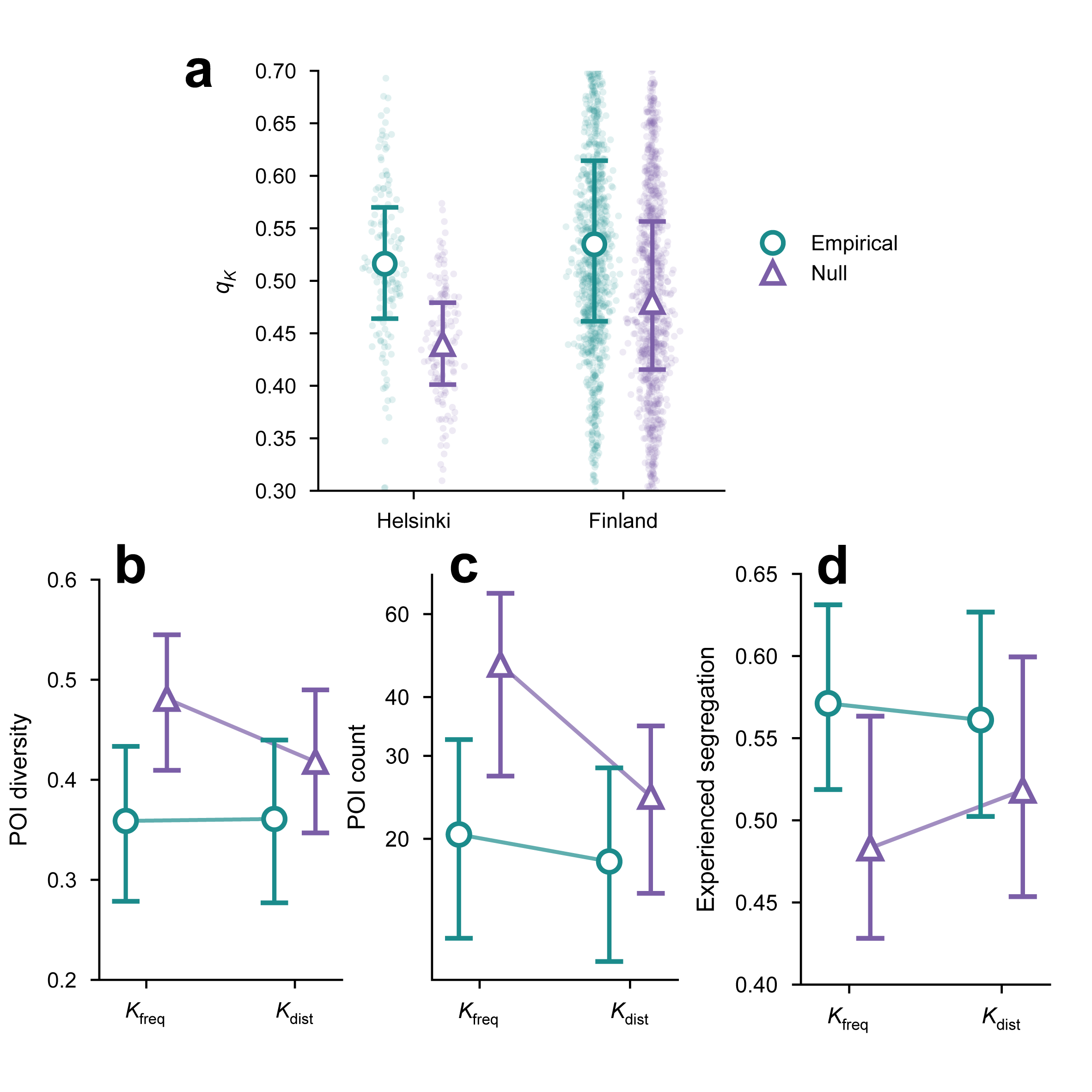}
  \caption{\textbf{Empirical mobility differs from the density-seeking null model.}
  \textbf{a.} Mobility alignment (${q_K}$) in the Helsinki region and nationwide for empirical mobility (teal circles) and the d-EPR null model (purple triangles). Empirical alignment is consistently higher, indicating that individuals remain more locally anchored than a purely density-seeking process predicts.
  \textbf{b--d.} Social and functional composition of recurrent ($K_{freq}$) and nearest ($K_{dist}$) visitations under the empirical and null scenarios: \textbf{b.} POI diversity, \textbf{c.} POI count and \textbf{d.} experienced segregation. The null model produces recurrent destinations with more POIs, greater diversity and lower segregation than observed empirically. Markers show means and error bars span the 25th--75th percentiles.}
  \label{fig:null}
\end{figure}

Empirical $q_K$ is higher than $q_K^{null}$ in both the Helsinki region and Finland (Fig.~\ref{fig:null}a): residents bypass nearby opportunities, but remain more locally anchored than the structural attraction of density would predict.
A mobility process governed by density attraction and distance decay alone would therefore pull individuals further from their nearest options than observed. 
The low alignment identified above is consequently not a search for dense destinations. Accessibility may release mobility from proximity, but it does not determine where that mobility is directed.

Destination composition reinforces this distinction. Compared with empirical $K_{freq}$ destinations, those generated by the null model contained more POIs, greater functional diversity and lower experienced segregation (Fig.~\ref{fig:null}b--d). Empirical behaviour was therefore directed towards less functionally diverse and more socially sorted destinations than a density-seeking process would produce. 
Density explains part of the opportunity landscape as it directly provided choices for destinations, but not the destinations ultimately selected. The next section examines the functional and social attributes associated with this divergence.

\subsection*{Destination selectivity underlying overshoot}

Having established that recurrent visitation systematically departs from nearby alternatives, and that density alone cannot explain this divergence, we next examine the destinations involved. 
We define an \emph{overshooting destination} as a recurrent destination that is not included in the individual's nearest-location set, $K_{freq}\setminus K_{dist}$. Conversely, destinations belonging to both sets, $K_{freq}\cap K_{dist}$, are defined as aligned. 
We trained a gradient boosting classifier to distinguish overshooting from aligned destinations using destination attributes, residential location attributes and neighbourhood socioeconomic context. 
To prevent observations from the same residential area entering both training and validation sets, we used five-fold cross-validation grouped by PAAVO postal area. 
The predictions achieved a ROC-AUC of 0.884 and a macro-F1 of 0.796, demonstrating that overshooting and aligned destinations have distinguishable attribute profiles. We interpret these predictive associations using SHAP values \autocite{shap}.

\begin{figure}
    \centering
    \includegraphics[width=0.9\linewidth]{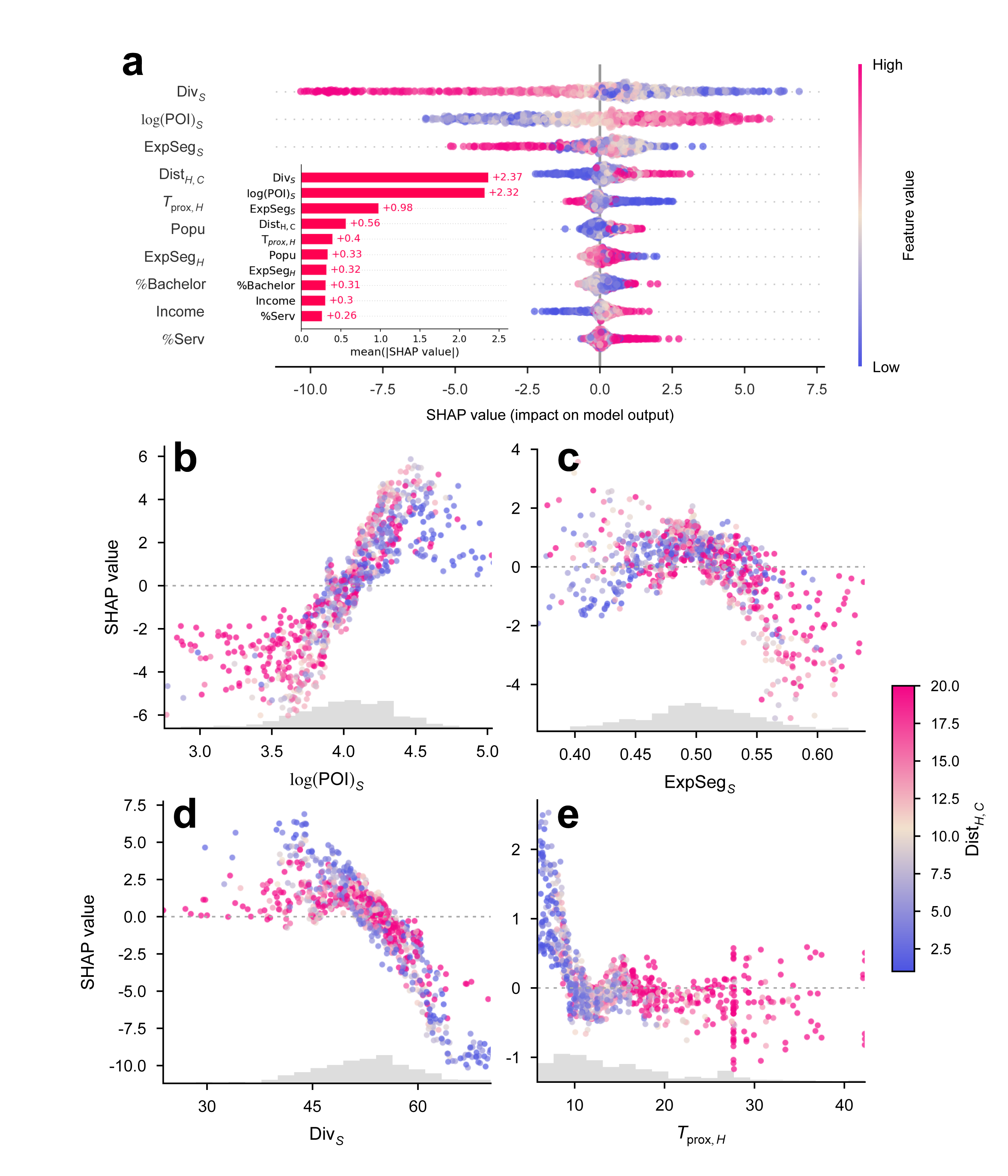}
    \caption{\textbf{Destination characteristics are the strongest predictors of overshoot.}
    \textbf{a.} SHAP summary plot for the gradient boosting classifier distinguishing recurrent non-nearest destinations (${K_{freq}\setminus K_{dist}}$) from recurrent destinations that are also proximate (${K_{freq}\cap K_{dist}}$). Positive SHAP values indicate contributions towards overshoot, and colours indicate feature values. The inset bars show mean absolute SHAP values for the ten most important predictors.
    \textbf{b--e.} SHAP dependence plots for log-transformed destination POI count ($\log(\mathrm{POI})_S$) \textbf{b.}, destination experienced segregation ($\mathrm{ExpSeg}_S$) \textbf{c.}, destination POI diversity ($\mathrm{Div}_S$) \textbf{d.}, and neighbourhood proximity travel time ($T_{\mathrm{prox},H}$) \textbf{e.}. Points are coloured by the residential neighbourhood's distance to the city centre ($\mathrm{Dist}_{H,C}$); grey histograms show feature densities.
    Higher POI counts increase the predicted likelihood of overshoot, whereas greater POI diversity generally reduces it, indicating that overshoot primarily targets amenity-rich but functionally specialised destinations.
    }
    \label{fig:shap}
\end{figure}

Destination attributes dominated the classification (Fig.~\ref{fig:shap}a). POI diversity, POI count and experienced segregation were substantially more influential than residential accessibility and socioeconomic characteristics. Overshoot was therefore distinguished primarily by where individuals travelled, while residential context provided secondary enabling conditions.
Overshooting destinations combined high POI counts with comparatively low functional diversity (Fig.~\ref{fig:shap}b,d). Recurrent mobility beyond nearest options therefore targeted amenity-rich destinations whose opportunities were concentrated within narrower functional portfolios.

The association with experienced segregation was nonlinear (Fig.~\ref{fig:shap}c). Intermediate levels contributed positively to overshoot predictions, whereas both low and high values were less consistently associated with overshoot. The results therefore indicate context-dependent social differentiation rather than a general tendency to select more segregated destinations. Exceptionally high residential accessibility, represented by very low $T_{prox}$, also contributed positively to overshoot predictions (Fig.~\ref{fig:shap}e), consistent with the earlier finding that strong accessibility can enable residents to move beyond their nearest options.

These destination-level associations help interpret the earlier divergence between recurrent visitation and nearest options. Exceptionally high residential accessibility is associated with recurrent travel beyond nearest options, consistent with accessibility enabling destination selectivity rather than enforcing local use. 
Residents of highly accessible central areas can therefore bypass nearby alternatives, sorting into destinations with particular functional and social profiles. 
In less accessible areas, destination selectivity may persist, but the greater travel cost of reaching alternatives can restrict its expression and contribute to stronger spatial alignment.

\subsection*{Amenity hierarchy structures recurrent visitation beyond one-size-fits-all proximity}
The 15mC implicitly treats amenity categories as similarly localisable under a common proximity threshold. Central Place Theory instead expects urban functions to differ by order and catchment: routine and frequently demanded services can often be distributed locally, whereas specialised, discretionary and infrastructural functions depend on wider demand, agglomeration and network connectivity \autocite{christaller1933zentralen,zhao2023investigating}. We therefore examine whether recurrent visitation differs from nearest options uniformly across amenity classes, or whether these differentials follow an amenity-specific hierarchy.

For each amenity class, we compare the mean distance to the nearest available option, $d_{prox}$, with the relative distance differential of recurrent visitation, $\Delta d_{rel}=(d_{freq}-d_{prox})/d_{prox}$, where $d_{freq}$ is the mean distance to recurrent destinations exposing the same amenity class. The x-axis captures baseline localisability, while the y-axis captures the proportional differential from that baseline. This distinction is important because a high relative differential does not necessarily imply a large absolute travel burden when the nearest option is already very close.

\begin{figure}
  \centering
  \includegraphics[width=0.9\linewidth]{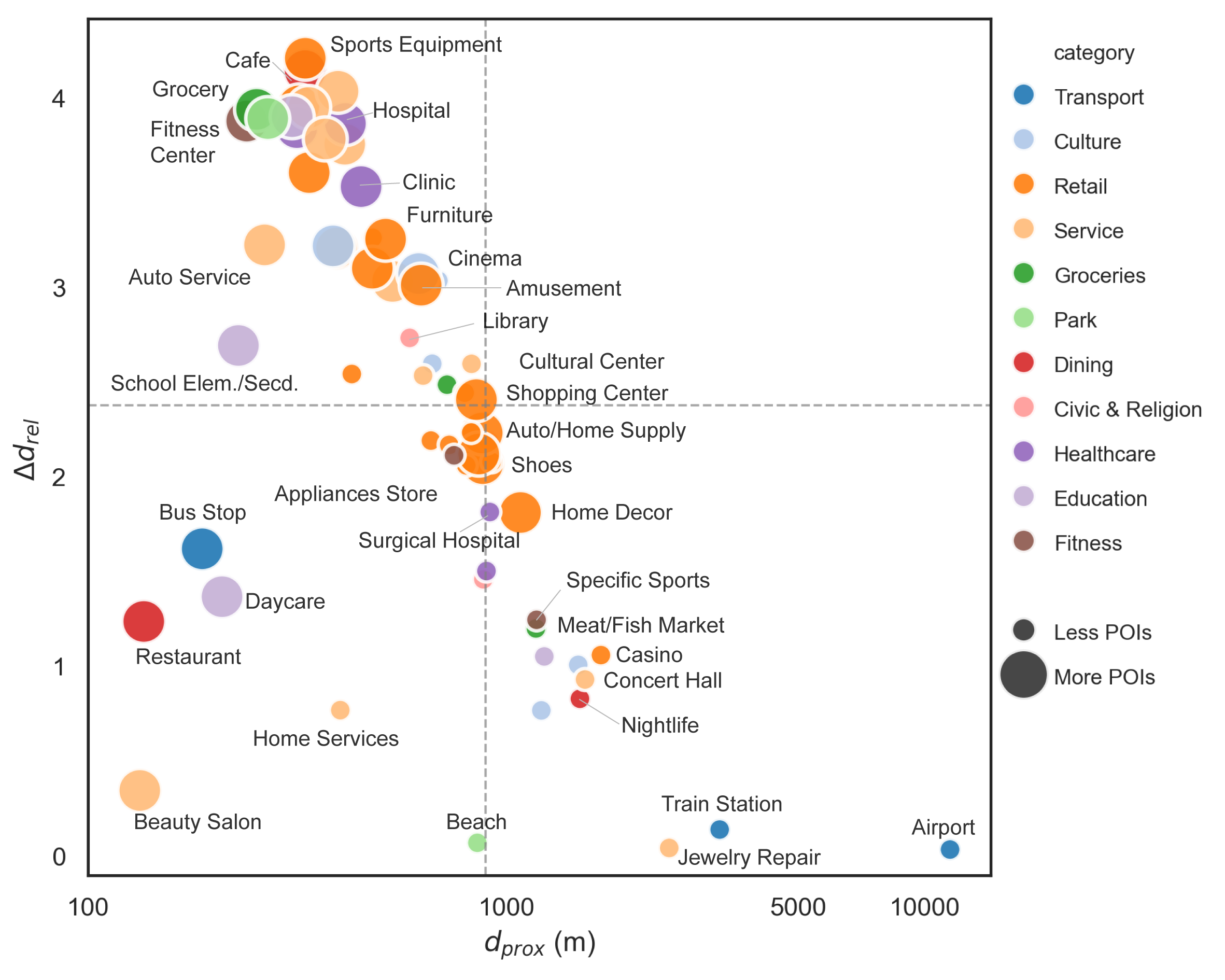}
  \caption{\textbf{Amenity-specific distance differentials follow a functional hierarchy.} 
  Relationship between the mean distance to the nearest available option, $d_{prox}$ (x-axis, log scale), and the relative distance differential of recurrent visitation, $\Delta d_{rel}=(d_{freq}-d_{prox})/d_{prox}$ (y-axis). Colours denote broad amenity categories and point size is proportional to the number of POIs in each class. The vertical and horizontal dashed lines represent mean values across amenity classes. Lower-left classes, such as bus stops, restaurants, daycare and schools, indicate routine anchors with short nearest-option distances and low relative differentials. Upper-left classes, such as groceries, cafes and fitness centres, indicate locally available amenities with stronger proportional differentials. Lower-right classes, such as airports and train stations, indicate structurally scarce amenities with long nearest-option distances but limited additional differentiation from the nearest option.
  }
  \label{fig:amenity}
\end{figure}

Three patterns emerge (Fig.~\ref{fig:amenity}). First, several routine anchors combine short nearest-option distances with low relative differentials. Bus stops, restaurants, daycare and schools fall into this regime, suggesting that recurrent visitation remains closely tied to nearby provision for these functions. These amenities are where the proximity logic of the 15mC is most behaviourally plausible. Second, some widely available amenities are locally available but still exhibit high proportional differentials. Groceries, cafes and fitness centres have short nearest-option distances, yet recurrent visitation often occurs at alternatives several times farther than the closest option. We interpret this as selective substitution among abundant nearby alternatives, or as the integration of these amenities into wider activity chains, e.g., coffee or groceries near work. Third, specialised and discretionary functions show a more conventional differentiation from nearest options. Cinemas, amusement facilities, furniture stores, shopping centres and cultural amenities occupy the upper or intermediate parts of the plot, indicating that recurrent visitation often favours more specific destinations over nearer alternatives. These functions are harder to reproduce evenly across neighbourhoods and depend more strongly on agglomeration and wider catchments.

The empty upper-right quadrant is also informative: no amenity class combines long nearest-option distances with high proportional differentials. Instead, unique infrastructure such as airports and train stations has long nearest-option distances but low relative differentials. This pattern reflects structural scarcity, because local substitutes do not exist and the nearest option already defines the feasible catchment. Together, the hierarchy is consistent with Central Place Theory: lower-order functions are most amenable to neighbourhood provision, higher-order and infrastructural functions are organised through wider catchments, and common amenities in dense supply create choice but do not automatically convert to nearest-option use. Proximity is therefore not equally behaviourally relevant across amenity types; implementation of the 15mC should secure routine access locally while maintaining sustainable connectivity for higher-order functions.

\section*{Discussion}

The 15mC has emerged as a compelling vision for sustainable urban living and a central planning paradigm. However, it often assumes that local provision will translate into local use, and it is frequently operationalised through a one-size-fits-all proximity target, i.e., the 15-minute neighbourhood. Recent scholarship has shown persistent gaps between local accessibility and realised mobility behaviour \autocite{maciejewska2025when, abbiasov202415minute}.
Using large-scale mobility data, we apply the K-Visitation framework to examine how realised behaviour, in the form of recurrent visitation, diverges from the proximity-based expectation, and what this divergence implies for the implementation of the 15mC.

A central insight from our analysis is that \textbf{proximity is a necessary but insufficient condition for local living}. Proximity shapes local accessibility and establishes a mobility baseline from which recurrent visitation often diverges, as individuals travel beyond the nearest available options. Recurrent visitation reveals that residents often spend extra travel time beyond what nearby opportunities would require.
This gap between recurrent visitation and proximity-based expectations is persistent and context-dependent. It is especially significant in dense, well-connected urban cores, where the physical conditions for proximity-based living are strongest but recurrent destinations depart most from nearest options both spatially and temporally. In these areas, dense amenity supply and strong connectivity reduce the marginal cost of bypassing nearest options, allowing residents to exercise destination selectivity. In peripheral areas, higher alignment may reflect constrained localism alongside longer travel burdens.

Our analysis further reveals that the divergence from nearest options reflects selective behaviour. The density-driven null model separates this selectivity from a pure amenity-density explanation: empirical recurrent visitation remains more locally anchored than a density-driven process would predict. By travelling beyond nearby options, residents target destinations with distinct functional and social profiles. These destinations are often amenity-rich but functionally specialised. Therefore, a generalised increase in amenity density does not guarantee local recurrent use, and densification should be based on behavioural evidence of preferences. This contrast between recurrent visitations and proximate options further quantifies an amenity hierarchy. The evidence challenges the notion that the same localisation target could be applied to different amenities. 

The results are especially relevant to the implementation goal of the 15mC. The policy value of the 15mC lies in securing local accessibility. Proximity provides a baseline condition for local living. A universal temporal threshold has limited validity as a target for realised individual behaviour. Behavioural evidence is a promising tool to identify where proximity is sufficient, where it is bypassed, and which functions require wider connectivity.
Furthermore, local implementation of the 15mC should be differentiated by amenity function and spatial context. Routine amenities, which are more compatible with local provision, should form the core of 15mC planning. Specialised and infrastructural functions depend on wider catchments and connectivity, and therefore require sustainable links across the metropolitan area.
The practical aim for planning is to secure local access to routine needs, reduce travel time burdens where proximity is weak, and maintain connectivity to functions that cannot be replicated in every neighbourhood.

Several limitations of this work should be acknowledged. 
While we have rigorously validated the mobile phone data, potential biases may persist due to inherent uncertainties in the collection algorithms, such as accuracy, irregularities, and completeness in the timing and frequency of location pings \autocite{kwan2018algorithmic, barreras2024exciting}. Our focus on residential behaviour excludes transient populations, such as tourists and intercity commuters, whose distinct non-local mobility patterns warrant future investigation.
Furthermore, inferred stay locations do not guarantee actual visits to specific amenities. For instance, a stop in an area with both a hospital and a bus stop POI does not confirm which, if either, was utilised \autocite{gong2014deriving}. This also applies to the experienced segregation measure, as we are not able to confirm social interactions from co-visitation \autocite{moro2021mobility}. Consequently, in our study, the amenity and segregation metrics are regarded as exposure measures, with visitations providing opportunities for visits or interactions.
The study is situated in Finland, a context characterised by lower density, relatively high equity, and well-developed sustainable transport infrastructure in larger cities. While some of our findings resonate with recent evidence from the US \autocite{abbiasov202415minute}, the social implications of proximity alignment may differ in hyper-dense contexts, such as Asian megacities or regions with different infrastructure profiles.
Finally, the K-Visitation framework currently weighs all amenity categories equally. Future research could refine this by integrating travel surveys to calibrate amenity weights based on specific demographic needs.

\section*{Methods}
\paragraph{Stay location data.} 

This study draws on anonymised mobile phone location data from Finland, covering the period from May 2023 to October 2024. The data were collected through a set of mobile applications in which users explicitly consented to share their location information. Raw location pings were grouped into stay locations using the \textit{InfoStop} algorithm \autocite{aslak2020infostop}, which clusters observations based on dwell time and spatial proximity. Following standard practice in the literature, we retained dwell times between 5~min and 24~h and aggregated pings within a 50~m radius. This procedure yielded approximately 104 million stay coordinates for around 720 thousand unique users.
Stay locations were aggregated to the H3 spatial hexagonal grid system at Level 9 resolution (approx. 174~m in side length) \autocite{h3}, which served as the basic unit of visitation.

Individual home and work-time locations were inferred based on their stay patterns. Home locations were identified as the most probable H3 grid where a user stayed between 23:00 and 07:00, while work-time locations were assigned as the most likely H3 grid visited between 10:00 and 16:00 on weekdays. 

Users without identifiable home or work-time locations in any month were excluded. 
We also excluded users present for fewer than 7 days and removed inter-region visits to filter out transient visitors. 
After filtering, the dataset comprised 11.9 million unique user-H3 grid combinations.
Stay-location processing, validation and transient-user sensitivity checks are shown in Supplementary Figs.~S2--S4 and S7.

For integration with socioeconomic attributes and more efficient visualisation, we further link users' home H3 grid to Finland's official national census grid with $1~\mathrm{km}\times1~\mathrm{km}$ cells \autocite{tilastokeskusstatistics}, and use it as our basis for visualisation. We tested the effect of changing aggregation sizes for potential MAUP issues, and the national grid provided robust results for our main metrics, $q_K$, while effectively preserving user data (Supplementary Fig.~S8).

\paragraph{Amenity and social exposure through visitation.}

To characterise the functional and social environment of each visitation, we integrate point-of-interest (POI) data as amenities and an experienced income segregation metric as a measure of social exposure. 
We acknowledge that mobile phone location data lack the spatial precision to distinguish entry into specific adjacent amenities; consequently, we do not infer verified ``check-ins,'' but instead treat visitation as \emph{exposure}--the set of functional and social opportunities available at a stay.
Amenities were assigned to each stay location by aggregating the POIs that fall within its H3 Level 9 hexagon, so that each visited cell directly defines its surrounding amenity environment.
The POI taxonomy was developed based on a synthesis of existing 15mC literature \autocite{moreno2021introducing,abbiasov202415minute,papadopoulos2023measuring, bruno2024universal} and includes ten categories: Civic \& religion, Culture, Dining, Education, Fitness, Groceries, Healthcare, Transport, Retail, and Services.

\textit{POI diversity.} We measure the diversity of amenity categories at a set of visitations using Shannon's entropy, $Div = -\sum_{c=1}^{10} p_c \ln p_c$, where $p_c$ is the proportion of POIs belonging to category $c$.

\textit{Experienced income segregation.} To quantify social exposure, we adopt the experienced income segregation metric of \autocite{moro2021mobility}. Each user is assigned a city-specific income quartile $q \in \{1,2,3,4\}$ based on the median income of their inferred home postal code area (PAAVO database, \autocite{2024paavo}). For each stay location $p$, the visitation share contributed by quartile $q$ is
\begin{equation}
    \tau_{pq} = \frac{\text{Total visits at } p \text{ by users in quartile } q}{\text{Total visits at } p \text{ by all users}},
    \label{tau}
\end{equation}
and the location's segregation score is its absolute deviation from perfect mixing (where $\tau_{pq} = 1/4$ for all $q$):
\begin{equation}
    s_p = \frac{2}{3}\sum_{q=1}^{4}\left|\tau_{pq} - \frac{1}{4}\right|.
    \label{S_p}
\end{equation}
An individual's experienced segregation, $S_i$, is the weighted sum of $s_p$ across their visitations.

\paragraph{Travel time by public transport.}
For each individual, travel time to their K-visitations was estimated using R5Py \autocite{fink2022r5py}, an open-source multimodal routing package, drawing on OpenStreetMap network data and GTFS public transport schedules from \textit{Traficom}. Following \autocite{ponkanen2025spatial}, we standardised key assumptions for Finland (departure times, walking speeds) and computed centroid-to-centroid public transport travel times between each user's home H3 cell and the relevant K-visitation cells.

From these estimates we derive three travel time quantities.
At the individual level, $T_{freq}$ and $T_{dist}$ summarise how a user's realised travel compares with the nearest options drawn from their own visited repertoire: 
(1) $T_{freq}$ is the frequency-weighted mean public transport travel time to the user's $K_{freq}$ cells, which reflects the user's realised access; and (2) $T_{dist}$ is the individual-level baseline, as the average public transport travel time to the user's $K_{dist}$ cells. It represents the nearest equivalents among the places that individual has actually visited.

At the home-cell level, we use $T_{prox}$ to provide a structural proximity benchmark based on all available opportunities, independent of individual mobility. We rank all reachable destination cells by public transport travel time, identify the smallest set covering all ten amenity categories, and define $T_{prox}$ as their mean public transport travel time. This shared benchmark indicates whether individuals travel beyond what their residential environment could support, rather than merely beyond their nearest visited options ($T_{dist}$).
\paragraph{d-EPR null model.}
To isolate the influence of urban structure from individual behavioural selectivity, we constructed a generative null model based on the density-aware Exploration and Preferential Return (d-EPR) framework \autocite{pappalardo2016human}.

This modelling framework generates synthetic visitation trajectories that preserve the fundamental statistical laws of human mobility--specifically, the probability of exploring new locations and the tendency to return to familiar ones--while replacing specific destination preferences with a probabilistic attraction to amenity density.

For each user $u$, we simulate a trajectory starting at their inferred home location $h_u$ with a length $T$ equal to their total observed displacements. At each step, the agent chooses to explore a new location with probability $P_{new} = \rho S_{loc}^{-\gamma}$, where $S_{loc}$ is the number of previously visited locations. We fit $\rho$ individually for each user (mean $\rho \approx 0.65$) and set $\gamma=0.23$ following established literature \autocite{song2010modelling,pappalardo2015returners}.

If exploring, the agent selects a destination $j$ from the H3 grid lattice based on a gravity-like competition between opportunity density and travel cost:
\begin{equation}
    \Pi_{ij} \propto \frac{N_j}{r_{ij}^\beta},
\end{equation}
where $N_j$ is the normalised POI density of grid $j$, $r_{ij}$ is the Euclidean distance from the current location $i$, and $\beta=2$ represents the friction of distance. 
If returning (probability $1-P_{new}$), the agent selects a location from their visited set $S_{loc}$ proportional to its visitation frequency. 
For more details about the model fitting and validation, please refer to Section~S4 in the Supplementary Materials.

We process the resulting synthetic visitations using the K-Visitation framework to generate comparable results with the empirical data. We derive a synthetic set of recurrent destinations $K_{freq}^{null}$ and nearest visitations (based on original home location) $K_{dist}^{null}$, and compute the null alignment coefficient $q_K^{null}$. $q_K^{null}$ quantifies the alignment expected if residents were purely maximising density access and minimising distance, without specific functional targeting. Comparing patterns of the empirical $q_K$ and $q_K^{null}$ allows for rejecting or accepting the null hypothesis.

\paragraph{Gradient boosting classifier for predicting overshoots.}
To analyse overshooting behaviour, we adopted the SHAP package \autocite{shap} with a gradient boosting classifier (XGBoost) to examine its correlated factors in urban environments. SHAP is a game-theory-based approach to explain a given machine learning model \autocite{lundberg2020local}. Tree-based predictive models, such as XGBoost, are very powerful and can capture non-linear correlations between variables \autocite{chen2016xgboost}.

We build the model as a binary classifier, treating the predicted output as the recurrent but not nearest available destinations ($K_{freq} \setminus K_{dist}$). It is labelled against the recurrent destinations that also correspond to the nearest option ($K_{freq} \cap K_{dist}$). This setup allows us to identify the predictors for overshoot.
We selected four input domains as predictors: socioeconomic characteristics of the home location; attributes of the stay location, including POI diversity and experienced income segregation; differences between these home and stay characteristics, capturing relative disparities; and estimated travel time between home and stay locations.

Details about variable selection and engineering, model training, and tuning can be found in Supplementary Section~S5. The tuned model achieved a ROC-AUC score of 0.884 and a macro-F1 of 0.796, substantially outperforming baseline models.

Given this well-trained model, we use the SHAP package to understand how input factors correlate with overshoot destinations. The quantification is shown as a SHAP value (Fig.~\ref{fig:shap}a--e). A large absolute SHAP value implies that the output is more sensitive to the given factor value in the model. In the summary plot (Fig.~\ref{fig:shap}a), the x-axis corresponds to the SHAP values, where we can observe different predictive contributions to the overshoot behaviour, while colours indicate higher and lower values of a feature.
In the dependence plot (Fig.~\ref{fig:shap}b--e), the y-axis corresponds to the SHAP value, while the x-axis indicates the feature's value. Each scatter point in the plot represents a sample, and the colouring indicates the value of the interacting feature. It is useful for detecting non-linear or interactive effects of different features on the prediction from the dependence plot \autocite{lundberg2020local}.
\paragraph{Data availability}
The mobile phone data used in this study are not publicly available. The POI dataset from Precisely is commercially available (https://www.precisely.com/product/precisely-places/). Transport feed data are available from Traficom (https://www.traficom.fi/fi), road network data from OpenStreetMap (https://www.openstreetmap.org/), and socioeconomic data from Statistics Finland (https://stat.fi).

\paragraph{Code availability}
All analysis was conducted in Python. Code for reproducing the results is publicly available at GitHub (https://github.com/xnzhang-33/k-visitation-finland). 

\printbibliography

\end{refsection}
\paragraph{Acknowledgements}
X.Z. and E.A. acknowledge the support of the Economic and Social Research Council (ESRC) under Grant No. ES/Y00180X/1. H.T. acknowledges the support of the Research Council of Finland under Grant No. 354342. We thank Andrew Renninger for many fruitful discussions.

\paragraph{Author contributions}
X.Z. and E.A. conceived the project and designed the experiments; A.P. and H.T. collected the data; X.Z. processed and analysed the data, carried out the experiments, and wrote the manuscript. All authors contributed to editing the manuscript.
\\
\paragraph{Competing interests}
The authors declare no competing interests.
\end{document}